\documentclass[a4paper,english,11pt]{article}
\usepackage[dvips]{graphicx,epsfig,color}
\usepackage{wrapfig,rotating}
\usepackage{amssymb,amsmath,array}
\usepackage{subfigure}
\usepackage{wrapfig}
\usepackage{sidecap}
\usepackage{geometry}

\begin{document}
\title{Predictions for pp single diffractive cross section at LHC}

\author{{M.G. Poghosyan}\\[1ex]
 CERN, CH-1211 Geneva 23, Switzerland
}
\date{}
\maketitle

A model based on Gribov’s Regge calculus was developed \cite{diff} and was
proposed to describe diffractive processes. 
In this note we present numerical vales obtained from [1] for the dependence of single diffraction cross-section on diffractive mass at $\sqrt{s} = 0.9, 2.76$ and 7 TeV.

\begin{table}[h!]
\begin{center}
\begin{tabular}{c|ccc}
mass   & 
\multicolumn{3}{c}{ $\Delta \sigma / \Delta M$ (mb$\cdot$GeV$^{-1}$)}
\\ 
interval (GeV) &  0.9 TeV  & 2.76 TeV & 7 TeV \\
\hline
  1.08 -    2  & 2.3843 &  2.7240 &  3.0111 \\
  2 -    3  & 1.1321 &  1.2968 &  1.4426 \\
  3 -    4  & 0.6762 &  0.7734 &  0.8627 \\
  4 -    5  & 0.4628 &  0.5277 &  0.5891 \\
  5 -    6  & 0.3435 &  0.3900 &  0.4354 \\
  6 -    7  & 0.2689 &  0.3039 &  0.3391 \\
  7 -    8  & 0.2186 &  0.2459 &  0.2741 \\
  8 -    9  & 0.1829 &  0.2047 &  0.2279 \\
  9 -   10  & 0.1564 &  0.1741 &  0.1936 \\
 10 -   20  & 0.0897 &  0.0974 &  0.1075 \\
 20 -   30  & 0.0454 &  0.0468 &  0.0508 \\
 30 -   40  & 0.0307 &  0.0302 &  0.0322 \\
 40 -   50  & 0.0236 &  0.0222 &  0.0233 \\
 50 -   60  & 0.0194 &  0.0176 &  0.0181 \\
 60 -   70  & 0.0167 &  0.0146 &  0.0148 \\
 70 -   80  & 0.0149 &  0.0125 &  0.0125 \\
 80 -   90  & 0.0135 &  0.0110 &  0.0108 \\
 90 -  100  & 0.0124 &  0.0098 &  0.0096 \\
100 -  200  & 0.0099 &  0.0067 &  0.0061 \\
\end{tabular}
\end{center}
\end{table}


\begin{thebibliography}{9}
\bibitem{diff} A.B. Kaidalov and M.G. Poghosyan, in Proceedings of
the 13th International Conference On Elastic and Diffractive Scattering
(blois Workshop) 09. ArXiv:0909.5156 [hep-ph].
\end{thebibliography}
\end{document}